\documentstyle[pra,aps,epsf]{revtex}
\begin{document}
\draft

\title{\bf Exact boundary conditions at finite distance
for the time-dependent Schr\"odinger equation}

\author{M. Mangin-Brinet, J. Carbonell, C. Gignoux}
\address{Institut des Sciences Nucl\'eaires\\
        53, Avenue des Martyrs\\
        38026 Grenoble, France}
\date{\today}
\maketitle

\begin{abstract}
Exact boundary conditions at finite distance for the
solutions of the time-dependent Schrodinger equation are derived.
A numerical scheme based on Crank-Nicholson method is proposed 
to illustrate its applicability in several examples.
\end{abstract}

\pacs{: 03.65.Bz, 03.65.Nk, 03.80.+r}

\section{Introduction}

The time-dependent quantum mechanics formalism has been  extensively used in the recent years
in several problems of physics and chemistry.
Since the work of McCullough and Wyatt \cite{MW 71}
 showing the interest of dealing directly with non-stationary states
in calculating reactions probabilities,
a consequent number of works has been devoted to the wave-packet dynamics.
The reader could find the more recent developments in the reviews 
\cite{TDQMD 92,KLEBER 94,GS 95,BKS 97} and references therein.
 
When solving the time-dependent Schr\"odinger equation in a finite volume,  
one is faced with the problem of implementing the proper boundary conditions at finite distance.
These boundary conditions are usually those of a free wave outgoing propagation.
For the stationary solutions they are given by
fixing the value of the logarithmic normal derivative on the boundaries of the interaction domain. 
However for the case of non-stationary wave packets a satisfactory solution is not known.

The usual way to overcome this difficulty is to impose the nullity of the wave on the boundaries and push them
far enough to avoid the effect of parasite reflections in the region of interest \cite{ASS 67}.
Another way, widely used in other branches of physics and mechanics as well,
is by means of the so-called absorbing boundary conditions \cite{LW 78,KK 86,VB 92}.
They indeed minimize parasit reflections but perturbe the dynamics near the boundary
and do not provide a solution of the initial equation in its neighborhood.

A third way is by the so called wave function splitting 
algorithm which consists in removing from the total wavefunction
the part localized outside the interaction region before it reaches the boundary \cite{HM 87}.
Other possibilities based on the interaction picture formalism
have also been developed \cite{ZHANG 89}.

Although the above quoted methods provide a practical way to solve 
the problem they do not give a theoretical solution of it.
The aim of this paper is to formulate an exact boundary condition (EBC)
at finite distance for outgoing solutions of the time-dependent Schr\"odinger equation.
Consequently, it would allow a solution of time-dependent quantum mechanical problems 
in a finite spatial domain without any approximation in the boundaries. 
This theoretical result is completed by implementing the derived boundary conditions
in some practical calculations concerning the more usual problems of the wavepacket dynamics.

The paper is organized as follows: in Section 2 we derive the general expression 
for the EBC which has to be imposed at the
boundary of an arbitrary integration domain containing the interaction.
This conditions turns to be non local on time and
allows a solution limited to the interaction domain of the time-dependent Schr\"odinger equation. 
In Section 3 the boundary conditions are reformulated for a discrete-time evolution 
problem in the frame of the Crank-Nicholson scheme. 
Section 4 is devoted to the one dimensional Schr\"odinger equation for which the use of these 
conditions is particularly simple. 
Several illustrative examples are given in 
Section 5: the spreading of a free wave
packet, the scattering of a wavepacket on a static potential,
the behavior of an initially localized wave packet under a time-dependent perturbation.
In the last example we derive explicit solutions of the Schr\"odinger equation in a time-dependent delta potential.
The applicability of the formulated conditions in a two-dimensional problem is shown in Section 6.
A discussion on the interest and limitations of the EBC is finally presented in the conclusion.

\section{Boundary conditions at finite distance for the time-dependent Schr\"odinger equation}

Let us consider the dimensionless time-dependent Schr\"odinger equation in a d-dimensional configuration space
\begin{equation}\label{CSE}
\left[i\partial_t+\nabla_{\vec{x}}^2-V(\vec{x},t) \right]\Psi(\vec{x},t)=0
\end{equation}
The potential $V$, eventually time-dependent,
is assumed to vanish outside a finite domain $D_I$ named the interaction domain. 

\begin{figure}[hbtp]
\begin{center}\epsfxsize=10cm\mbox{\epsffile{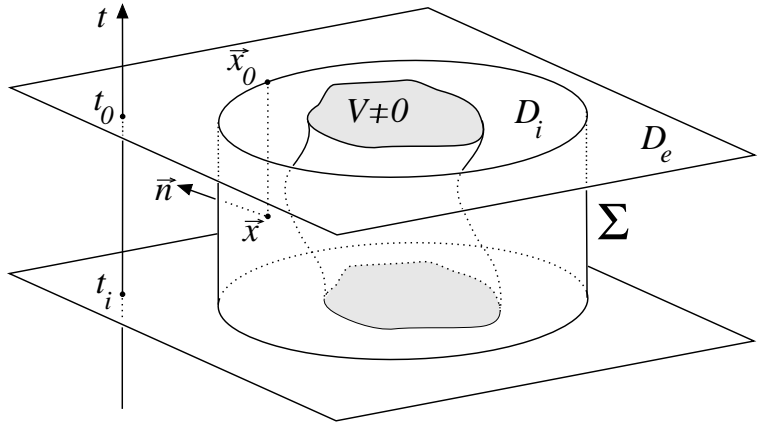}}\end{center}

\vspace{+1.5cm}
\caption{Space-time representation of different equal-time sections of the configuration space D.
The shadowed region represents the interaction domain $D_I$. 
The time-dependent Schrodinger equation is solved in the finite domain $D_i$}\label{fig1}
\end{figure}

We wish to obtain a solution of (\ref{CSE}) by solving the
equation in a finite spatial region $D_i$, called hereafter the integration domain.
The domain $D_i$ has to be chosen such that it contains $D_I$, i.e.
such that its complementary $D_e$ with respect to the total configuration space $D$ is free of interaction. 
We have thus $D=D_i\bigcup D_e$ and $D_I\subseteq D_i$ ( see figure \ref{fig1}).

The solution of (\ref{CSE}) in $D_i$ requires to formulate and implement the proper boundary 
conditions on its border, a surface $\Sigma$.
For this purpose 
let us first consider the solution $\Psi$ of (\ref{CSE}) in $D_e$.
The domain $D_e$ is, by definition, free of interaction and $\Psi$ satisfies the free Schr\"odinger equation  
\begin{eqnarray} 
(i \partial_{t} + \nabla_{\vec{x}}^2) \Psi(\vec{x},t)&=&0 \label{FCSE}
\end{eqnarray} 
Let $K_0^+$  be the free retarded propagator, i.e. a solution of
\begin{eqnarray} 
(-i \partial_{t} + \nabla_{\vec{x}}^2) K_0^{+}(\vec{x_0}-\vec{x},t_0-t)
&=&i \delta(\vec{x_0}-\vec{x}) \delta(t_0-t) \label{112}
\end{eqnarray} 
with $\vec{x}\in D_e$ and $(\vec{x}_0,t_0)$ arbitrarily chosen.

We have in mind the outgoing solution at $(\vec{x}_0,t_0)$
of a time-dependent problem with initial value given by $\Psi(\vec{x},t_i)$ for $t_i<t_0$.
By multiplying equation (\ref{FCSE}) by $K_0^+ $, equation (\ref{112}) by $\Psi$ and substracting one gets the relation
\begin{eqnarray} \label{eq2} 
\Psi(\vec{x},t)\delta(\vec{x}_0-\vec{x})\delta(t_0-t)&=&-\partial_t\left[K_0^{+}(\vec{x}_0-\vec{x},t_0-t) \Psi(\vec{x},t)\right]\cr
&-&\nabla_{\vec{x}}\cdot\left[i\Psi(\vec{x},t)\stackrel{\leftrightarrow}{\nabla}_{\vec{x}}
 K_0^{+}(\vec{x}_0-\vec{x},t_0-t)\right]
\end{eqnarray}
with $f{\stackrel {\longleftrightarrow} {\nabla}} g=f(\vec{\nabla}g)-(\vec{\nabla}f)g$.

This relation is to be integrated over the $d+1$ dimensional space-time volume
$D_e\times[t_i,+\infty]$. We will assume that the domain $D_e$ is connected. This assumption is justified
for most of the cases except for $d = 1$, which will be discussed later.    
The integral of the right hand side is then
transformed, by using the Green theorem, into a flux term through the boundary surface $\Sigma$.
We denote by  $\vec{n}$ the normal to $\Sigma$ pointing towards $D_e$ (see figure \ref{fig1}).  

Taking into account that $K_0^{+}(\vec{x_0}-\vec{x},t_0-t)=0$  for t$>t_0$ and
that there is no contribution  from the flux at the infinity for purely outgoing waves (for
they have the same asymptotic behavior as the retarded free propagator)
we obtain the following relations depending on the relative position of $\vec{x}_0$.

If $\vec{x}_0\in D_e$ one has
\begin{eqnarray} \label{BCDE}
\Psi(\vec{x_0},t_0)=                                      
\int_{D_e} K_0^{+}(\vec{x_0}-\vec{x},t_0-t_i)\Psi(\vec{x},t_i)d\vec{x}\\
+i\int_{t_i}^{t_0}dt\int_{\Sigma}\Psi(\vec{x},t){\stackrel {\longleftrightarrow} {\nabla}} K_0^{+}(\vec{x_0}-\vec{x},t_0-t).\vec{n}dS\nonumber 
\end{eqnarray}
The solution in the free region is thus obtained
as a sum of a first term, representing the free propagation of the 
wave which was initially in the non interacting region $D_e$,
and a second term made only with
the values of the wavefunction and its normal derivative at the boundary $\Sigma$ at the past times.
Equation (\ref{BCDE}) can be consequently used to propagate outside $D_i$ a solution obtained in $D_i$
and can be viewed as a generalization of the Huyghens principle in a space-time surface.

If $\vec{x}_0\in\Sigma$ one has
\begin{eqnarray} \label{BCDS}
{1\over2}\Psi(\vec{x_0},t_0)=                                      
\int_{D_e} K_0^{+}(\vec{x_0}-\vec{x},t_0-t_i)\Psi(\vec{x},t_i)d\vec{x}\\
+i\int_{t_i}^{t_0}dt\int_{\Sigma}\Psi(\vec{x},t){\stackrel {\longleftrightarrow} {\nabla}} K_0^{+}(\vec{x_0}-\vec{x},t_0-t).\vec{n}dS\nonumber 
\end{eqnarray}
This result can be obtained either as a limiting case of the preceding one (\ref{BCDE}) or by directly integrating
equation (\ref{eq2}) for $\vec{x}_0\in\Sigma$.
Equation (\ref{BCDS}) is the boundary condition we were looking for.
It provides an integral relation between the value of the wavefunction on a point of $\Sigma$ at a given time
and the values of the wavefunction and their derivatives in $\Sigma$ at the preceding times.
Starting from a known initial state at time $t_i$, equation (\ref{BCDS}) gives in its
differential form the boundary condition
on $\Sigma$ required for determining at time $t_i+\Delta t$ the solution
of the time-dependent Schrodinger equation in the finite domain $D_i$.

Several comments are in order:
\begin{itemize}

\item The relation (\ref{BCDS}) is non local in both space and time. It generalizes
the outgoing boundary condition for the monochromatic free plane waves, 
which in the one-dimensional case and at $x=a$ reads:
\begin{equation}\label{bcpw1d}
\Psi(a,t_0)={1\over ik}\partial_x\Psi(a,t_0)
\end{equation}

\item The choice of $K_0^+$ ensures that only free outgoing waves will propagate across $\Sigma$,
in the same way as a solution obtained in an infinite domain.
This result is exact and consequently  will not generate any reflection on the boundary.

\item The first term of its right hand side corresponds to the free propagation of the initial state 
which was initially in the external region $D_e$.
Its numerical calculation needs some care due to the oscillatory character of the propagator
and several ways to overcome it are proposed below:
\begin{itemize}
\item It is always possible to eliminate this contribution
by choosing the initial time $t_i$ and $D_i$ in such a way that the initial state is fully in $D_i$.

\item In the case when the initial state is very extended, it is more suitable to include it entirely in $D_e$. 
Indeed the volume integral term gives then the value at the boundary of a free evolution of the initial state.
This can be easily calculated in the momentum space
and is known analytically for some cases of physical interest like  gaussian wavepackets.

\item In the case when the initial state is entirely in $D_e$,
there exists another possibility to remove the volume term by splitting
 the solution into its non perturbed and scattered parts, i.e.
\[ \Psi=\Phi  + \chi \]
in which $\Phi$ is the solution of (\ref{FCSE}) which coincides at $t=t_i$ with the initial state 
and $\chi$ a solution of the inhomogeneous equation
\begin{equation}\label{ICSE}
\left[i\partial_t+\nabla_{\vec{x}}^2\right]\chi(\vec{x},t)=V(\vec{x},t)\Phi(\vec{x},t) 
\end{equation}
The unknown function $\chi$ vanishes at $t=t_i$
and satisfies in the free domain $D_e$ the same equations
than $\Psi$. It will consequently satisfy on $\Sigma$ the boundary conditions (\ref{BCDS}) 
without the volume term. 
\end{itemize} 
\end{itemize}

The preceding relations, completed by the case $\vec{x}_0\in D_i$, are summarized in the following equation
\begin{eqnarray} \label{OBC}
\epsilon\Psi(\vec{x_0},t_0)=                                      
\int_{D_e} K_0^{+}(\vec{x_0}-\vec{x},t_0-t_i)\Psi(\vec{x},t_i)d\vec{x}\\
+i\int_{t_i}^{t_0}dt\int_{\Sigma}\Psi(\vec{x},t){\stackrel {\longleftrightarrow} {\nabla}} K_0^{+}(\vec{x_0}-\vec{x},t_0-t).\vec{n}dS\nonumber 
\end{eqnarray}
with $\epsilon$  defined by
\[\epsilon(\vec{x}_0)=
\left\{\begin{array}{lll}
        1&{\rm if}&\vec{x}_0\in D_e \cr
        0&{\rm if}&\vec{x}_0\in D_i \cr
{1\over2}&{\rm if}&\vec{x}_0\in \Sigma \end{array}\right. \]

The above derivation has been done under the assumption that the potential
$V$ vanishes outside a finite domain $D_I$.
This condition can be however weakened in some cases.
For instance in the case of an interaction made of one short range part 
$V_s$ plus a static long range
part $V_L$ (e.g. Coulomb, polarisation or centrifugal potentials) the same derivation holds.
The interaction domain is then defined by $V_s$ whereas $D_e$ can contain $V_L$ all across.
The only difference in equation (\ref{OBC}) consists in replacing the free propagator $K_0^+$
by the corresponding propagator of the long range potential $V_L$.

In the one dimensional case the interaction domain is $D_i=[-a,a]$
and the free exterior domains $D_e$ has two connected components $[-\infty,-a]$ and $[a ,\infty]$.
The boundary condition (\ref{OBC}), based on the Green theorem, has to be applied to each of them. 
Let us take for instance $D^{+}_e=[a ,\infty]$. At $x_0=a$, the normal derivative
of the free propagator on the boundary  vanishes,
as it can be explicitely checked on its expression ($\hbar=2m=1$): 
\begin{equation}
K_0^{+}(a-x,t_0-t)=\Theta(t_0-t){1\over\sqrt{4i\pi(t_0-t)}} e^{ {i\over4}{(a-x)^2\over t_0-t} }
\end{equation}
The EBC (\ref{BCDS}) becomes simply 
\begin{equation}\label{OBC1} 
{1\over2}\Psi( a,t_0)=\int_{a}^{\infty} K_0^{+}(a-x,t_0-t_i) \Psi(x,t_i) dx
-\int_{t_i}^{t_0}dt\sqrt{\frac{i}{4\pi(t_0-t)}} \partial_x\Psi(x,t) \vert_{a}
\end{equation} 

It is worth noticing that in case $t_i=-\infty$ this result may be obtained
in a quite straightforward way.
Indeed let us consider the Fourier energy components of the wavefunction
\begin{equation}\label{OBC2} 
\Psi(x,t_0)=\int_{-\infty}^{+\infty} d\omega \Psi(x,\omega)e^{-i\omega t_0}
\end{equation} 
and let us impose to each of them the plane wave outgoing  boundary condition (\ref{bcpw1d}) on the form
\[ i\sqrt{\omega}\Psi(a,t_0)=\partial_x\Psi(a,t_0)\]
Inserting this relation in (\ref{OBC2}) one gets 
\begin{equation}\Psi(a,t_0)=\int_{-\infty}^{+\infty}dt\int_{-\infty}^{+\infty} 
d\omega \frac{e^{-i\omega(t_0-t)}}{2i\pi\sqrt{\omega}}
\partial_x\Psi(a,t)\end{equation}
We recognize the Fourier transform of the free Green function for zero argument 
\begin{equation}
K_0^+(x,\omega)=-i\int_{-\infty}^{+\infty} dt\;e^{-i \omega t} K_0^+(x,t)={1\over 2i\sqrt{\omega}}
e^{i\sqrt{\omega}|x|}\nonumber\end{equation}
and we obtain the following relation: 
\begin{equation}
\Psi(a,t_o)=-\int_{-\infty}^{t_0} dt \sqrt{ {i }\over {\pi(t_0-t)}}\partial_x\Psi(a,t) 
\end{equation}
in agreement with the second term in the right hand side of (\ref{OBC1}).
The first term does not contribute in the limit $t_i\rightarrow-\infty$
due to the fact that the free propagator tends to zero.

The boundary condition (\ref{OBC}) generalizes the KKR \cite{KKR} method which provides a relation 
between the solution and its normal derivative at the boundary of a free domain.
This method has been derived and successfully applied in the case of static quantum billards. 
The formulation we have developed in this section appears as a natural extension of this method
for space-time billards.

The condition (\ref{OBC}) is not directly useful in a numerical solution of the problem.
Indeed the singularity of the propagator at $t=t_0$ would generate
instabilities when the dependent Schr\"odinger equation is discretized. To overcome this difficulty
one has to reformulate the boundary conditions in the particular approximate scheme chosen.
The aim of the following section is to obtain such a formulation in the frame 
of the Crank-Nicholson scheme.

\section{Boundary conditions in the Crank-Nicholson scheme}

The Cranck-Nicholson \cite{CN} method for solving a discrete-time evolution problem
consists in approximating the infinitesimal evolution operator
\[ {\left| {\Psi(t+\Delta t)} \right\rangle}= e^{-i{H\over\hbar}\Delta t} {\left| \Psi(t) \right\rangle} \]  
by the expression:
\begin{equation}  \label{CN1}  
(1+{iH \over 2\hbar} \Delta t) {\left| \Psi_{m} \right\rangle}=(1-{iH \over 2\hbar}  \\  
 \Delta t) {\left| \Psi_{m-1} \right\rangle}  
\end{equation}
where ${\left|\Psi_{m}\right\rangle}$ and  ${\left|\Psi_{m-1}\right\rangle}$ are
the wavefunctions at two consecutive times separated by $\Delta t$ and
the hamiltonian H is evaluated at the mean time $t+{\Delta t\over2}$. 

By introducing the imaginary parameter $\mu^2=\frac{4im}{\hbar \Delta t}$, equation (\ref{CN1}) can be
rewritten in the form
\begin{equation}\label{DSE}
(\mu^2-H)\Psi_{m}=(\mu^2+H)\Psi_{m-1}
\end{equation}
according to what the solution at time-step m is obtained
by solving an inhomogeneous stationary Schr\"odinger equation with complex energy $\mu^2$.
The source term is provided by the solution at the preceding time.
  
By means of (\ref{DSE}) the equivalent of equation (\ref{FCSE}) for the outgoing
wavefunction in the free domain $D_e$ becomes 
\begin{equation}\label{FDSE}
\mu^2(\Psi_{m}-\Psi_{m-1})=-\nabla_{\vec{x}}^2(\Psi_{m}+\Psi_{m-1})
\end{equation}
and the discrete free retarded propagator $K_p(\vec{x_0}-\vec{x})$ is defined as the solution of
\begin{equation} \label{CN3}\mu^2( K_{p+1}-K_{p} )=
-\nabla_{\vec{x}}^2( K_{p+1}+K_{p} )\end{equation}
normalised to $K_0(\vec{x_0}-\vec{x})=\delta(\vec{x_0}-\vec{x})$ 

For $\vec{x}_0\notin\Sigma$ one obtains the analogous to the continuum case
by simply replacing the time integral by a sum over the discrete index p
\begin{eqnarray} \label{CN5} 
\epsilon\Psi_n(\vec{x_0})&=&\int_{D_e} K_n(\vec{x_0}-\vec{x})\Psi_0(\vec{x}) dV 
\\&+&
\frac{1}{2\mu^2}\sum\limits_{p=0}^{n-1}\int_{\Sigma}
\left[
      \left( K_p(\vec{x}-\vec{x}_0) +    K_{p+1}   (\vec{x}-\vec{x}_0) \right)
{\stackrel {\longleftrightarrow} {\nabla}}  \left( \Psi_{n-p}(\vec{x})    + \Psi_{n-p-1} (\vec{x}          ) \right)
\right]\vec{n}dS \nonumber
\end{eqnarray}
In calculating the limit $\vec{x_0}\rightarrow\Sigma$ and like in the continuum case
there is a singularity in the normal derivative of the free propagator.
Its contribution to the surface term in the r.h.s of (\ref{CN5}) is
\[\frac{1}{2}\sum\limits_{p=0}^{n-1}(-1)^p(\Psi_{n-p}(\vec{x_0})+\Psi_{n-p-1}(\vec{x_0})) 
=\frac{1}{2}\left(\Psi_{n}(\vec{x_0})+(-1)^n\Psi_{0}(\vec{x_0})\right)  \]
for $\vec{x_0}\in D_e$ and the opposite for $\vec{x_0}\in D_i$. 
The volume contribution can be in its turn re-written by means of the regularized 
propagator $\tilde{K}_p$ defined as
\begin{equation}\label{propa}
K_p(\vec{X})=(-1)^p\delta(\vec{X})+\tilde{K}_p(\vec{X})
\end{equation} 
and one finally arrives  to  
\begin{eqnarray} \label{DOBC} 
{1\over2}\Psi_n(\vec{x_0})&=&\int_{D_e}\tilde{K}_n(\vec{x_0}-\vec{x})\psi_0(\vec{x})dV 
\\&+&
\frac{1}{2\mu^2}\sum\limits_{p=0}^{n-1}\int_{\Sigma}
\left[ \left( \tilde{K}_p(\vec{x}-\vec{x_0}) +\tilde{K}_{p+1}(\vec{x}-\vec{x}_0) \right)
{\stackrel {\longleftrightarrow} {\nabla}}   \left( \Psi_{n-p}(\vec{x})+\Psi_{n-p-1}(\vec{x})                   \right)
\right]\vec{n}dS \nonumber
\end{eqnarray}
This expression is the  equivalent of (\ref{BCDS}) for the discrete time evolution.

To make use of the discrete EBC condition (\ref{DOBC}) it remains
to derive an expression for the discrete-time free propagator $K_p$.
In the momentum space equation (\ref{CN3}) reads as 
\[K_{p+1}(\vec{k})=\left(\frac{\mu^2+k^2}{\mu^2-k^2}\right) K_{p}(\vec{k})\]
which leads to: 
\begin{equation}  \label{as}
K_{p}(\vec{k})=\left(\frac{\mu^2+k^2}{\mu^2-k^2}\right)^p
\end{equation}
One can see in equation (\ref{as}) that in the limit $k\rightarrow\infty$ the Fourier transform of 
the propagator contains the singular delta function manifested in (\ref{propa}). 
Nevertheless this term cancels in the sum of two successive propagators.
This sum is given by the non singular integral:  
\begin{equation} \label{K+K}
K_{p}(\vec{X})+K_{p+1}(\vec{X})=
\int\frac{2\mu^2}{\mu^2-k^2}\left(\frac{\mu^2+k^2}{\mu^2-k^2}\right)^p
e^{i\vec{k}\vec{X}} \frac{d\vec{k}}{(2\pi)^d}
\end{equation}

This integral can in principle be calculated by the contour Cauchy method.
However, its value can also be obtained by means of the discrete Fourier tranform.
We note indeed that a N discrete-time Fourier transfom with respect to the index p of equation (\ref{K+K}) 
gives the sum of a geometric series. With the inverse Fourier transform we obtain:
\begin{equation} \label{Kdx}
K_{p}(x)+K_{p+1}(x)=\frac{1}{N}\sum\limits_{l=0}^{N-1}
\frac{e^{-ip\phi_l}}{1+e^{i\phi_l}}G_0^+(k_l^2,x)
\end{equation}
where $\phi_l=\frac{2l\pi}{N}+i\eta$ and $\eta$ is the required convergence factor of the geometric series 
which has to be chosen such that $e^{-N\eta}<<1$.
$G_0^+$ is the standard free Green function with energy $k_l^2=\mu^2\frac{1-e^{i\phi_l}}{1+e^{i\phi_l}}$.

Although the formalism has been developed in view of solving the Schrodinger
equation in a finite domain $D_i$, equation (\ref{CN5}) with $\epsilon=1$
provides the wavefunction in all the configuration space.
However one can obtain a closed form for some observables 
in terms of the internal solution only.
Let us consider for example the overlapping between two outgoing solutions
$\Phi$ et $\Psi$ of (\ref{FDSE}) in the exterior domain.
Standard algebra gives for the exterior part of the integral the expression
\begin{equation}\label{flux1}
{\left\langle\left.{\Phi}\right| {\Psi}\right\rangle} _e^n- {\left\langle\left.{\Phi} \right| {\Psi} \right\rangle}_e^{n-1}=
\frac{1}{2\mu^2}\int(\Phi_n + \Phi_{n-1})^*{\stackrel {\longleftrightarrow} {\nabla}}(\Psi_n+\Psi_{n-1}) \vec{n}dS
\end{equation}
which can be calculated from the solution in $D_i$.
By adding the corresponding equations for
consecutive time steps one obtains the desirate scalar product.
This expression is used e.g. in calculating the auto-correlation function 
(see section \ref{delta}).
The case $\Phi_n=\Psi_n$ provides the rate of variation of the probability
in the exterior domain by time step.
The contribution from $D_i$  can be calculated 
by numerical integration and we thus have
the possibility to verify the conservation of the total norm .

In summary the solution of the time dependent Schr\"odinger equation in the
interaction domain $D_i$ can be obtained by solving, at each time step, a stationary
complex and inhomogeneous Schr\"odinger equation (\ref{CN1}) in $D_i$ with 
the boundary conditions given by (\ref{DOBC}).
The local character of this equation is preserved 
by these conditions. We are thus let
with the solution of a banded linear system, whatever the discretisation algorithm
used for the spatial variables.
For static hamiltonians this linear system is in addition the
same at any time-step.

\section{Examples in the one dimensional case}

In the one dimensional problems, simplifications arise.
Like
in the continuum case, the normal derivatives of the propagator at the boundaries vanish.
The sum of two successive propagators for $x_0=a$ can be explicitly calculated and gives: 
\begin{equation} \label{K+K1}
K_{p}(0)+K_{p+1}(0)=
\int\frac{2\mu^2}{\mu^2-k^2}\left(\frac{\mu^2+k^2}{\mu^2-k^2}\right)^p
\frac{dk}{2\pi}
\end{equation}
This integral can be performed by standard methods: it vanishes for odd values of p and for $p=2q$ it
gives  
\begin{equation} \label{integ}
K_{2q}(0)+K_{2q+1}(0)=-i\mu C_q
\end{equation}
with
\[ C_q= \frac{(2q)!}{(2^q q!)^2}\]
The boundary conditions (\ref{DOBC}) give at $x_0=a$:
\begin{equation}\label{C1D}
\Psi_n(a)=2\int_{a}^{\infty}\tilde{K}_n(x-a)\Psi_0(x) dx -{i\over\mu}
\sum\limits_{q=0}^{[{n-1\over2}]}C_q\partial_x\left[\Psi_{n-2q}(a)+\Psi_{n-2q-1}
(a)\right]
\end{equation}
At $x_0=-a$  one obtains in a similar way
\begin{equation}\label{C1G}
\Psi_n(-a)=2\int^{-a}_{-\infty}\tilde{K}_n(x+a)\Psi_0(x) dx +{i\over\mu}
\sum\limits_{q=0}^{[{n-1\over 2}]}C_q\partial_x\left[\psi_{n-2q}(-a)+\Psi_{n-2q-1}(-a)\right]
\end{equation}

For illustration we have detailed some simple examples concerning the time evolution of wavepackets.
The initial state has been taken for simplicity of gaussian form
\begin{equation}\label{gwp}
\Psi_0(x)=\frac{1}{\pi^{1/4}\sigma_0^{1/2}}
e^{i{mv\over\hbar}(x-x_0)}\;e^{-\frac{(x-x_0)^2}{2{\sigma_0}^2}}
\end{equation}

The examples have been  solved by scaling the integration domain $D_i$ 
to the interval $[-1,+1]$. 
The discrete-time Schr\"odinger equation (\ref{CN1}) is there solved
at each of the N time-steps by finite differences
method with $N_x$ equally spaced grid points $x_i$.
The values of the wave function at $x_i$ are the unknowns.
The required $N_x$ equations are obtained by validating (\ref{CN1}) at each grid point.
However to evaluate the second derivative at the end of the grid the 
values of the wave function at the nearest points exterior of this domain is needed.
These are written in terms of the internal points
by using EBC conditions (\ref{C1D},\ref{C1G}) with a symmetric formula to evaluate the derivative term.
Substituting this exterior values in
(\ref{CN1}) we end with a tridiagonal linear system of dimension $N_x$.

\subsection{Free propagation of a gaussian wave packet}

The first example concerns the free propagation of a gaussian wave-packet.
Its main interest lies in the fact that an 
analytical solution is known which allows to check the validity of our boundary conditions.
In this case the integration domain is not fixed by the interaction
but chosen such that it fully contains the initial wavepacket.
 
The wave packet (\ref{gwp}) is initially centered ($x_0=0$) and,
in absence of interaction, the probability density at time $t$ is known to be given by :  
\[ \rho(x,t)=\mid\Psi(x,t)\mid^2 =
\frac{1}{\pi^{1/2}\sigma(t)} e^{-\frac{(x-vt)^2}{\sigma^2(t)}}\]
with $\sigma(t)=\sigma_0\sqrt{1+\tilde{t}^2}$ and $\tilde{t}=\frac{\hbar}{m\sigma_0^2}t$
a dimensionless time variable.
If the total evolution time in this units is $\tilde{T}$,
one has  $\mu^2=\frac{4im}{\hbar \Delta t}=\frac{4iN}{{\sigma_0}^2\tilde{T}}$.
We have taken for the initial width the value $\sigma_0=0.2$.

\begin{figure}[hbtp]
\begin{center}\epsfxsize=10cm\mbox{\epsffile{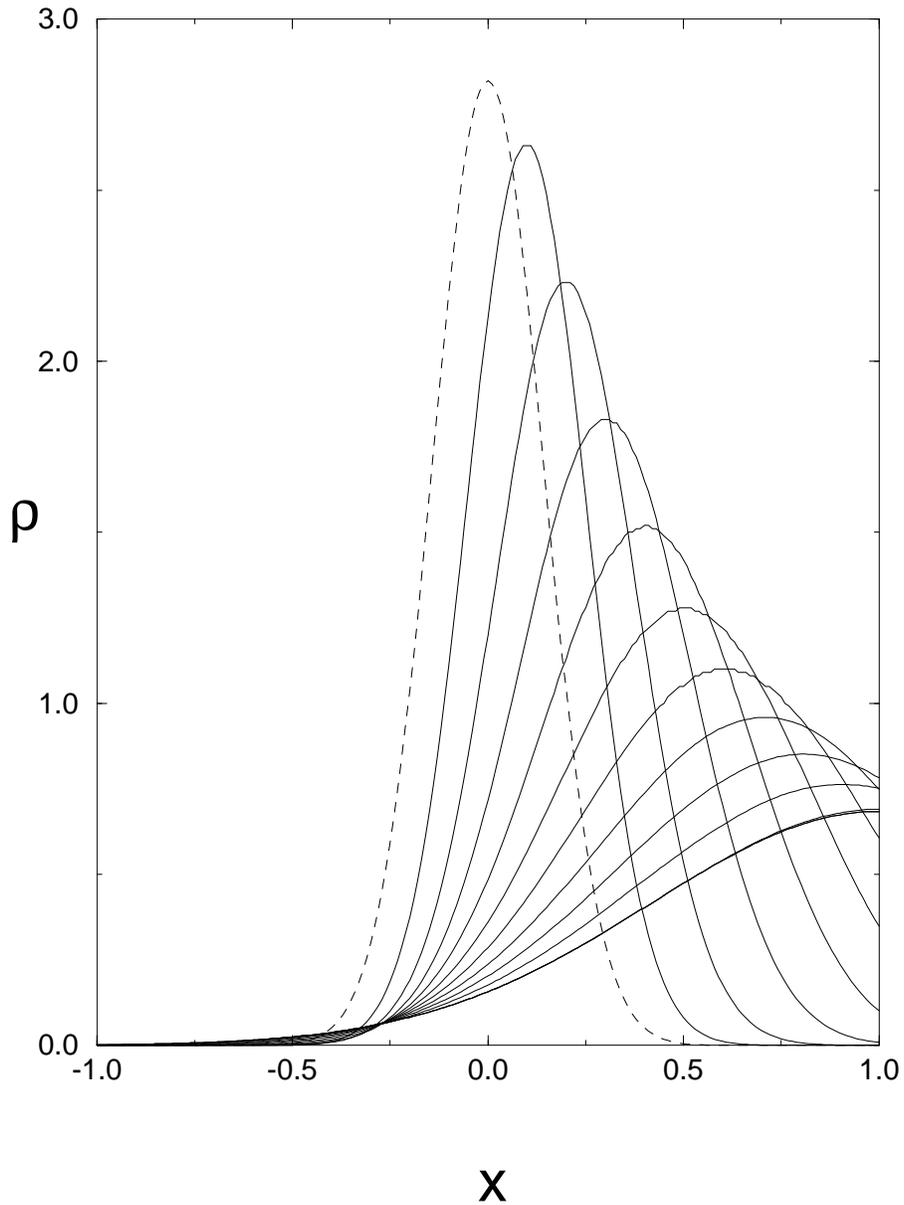}}\end{center}

\vspace{+3.5cm}
\caption{Free evolution of a gaussian wave packet.
It has been obtained by solving the time-dependent Schr\"odinger equation in a limited region 
of the configuration space with the  boundary conditions  given by (\protect{\ref{OBC}}).
Dashed curve represents the probability density of the initial wavepacket for $\sigma_0=0.2$.
The time evolution is plotted at ten different times showing the spreading of the wavepacket 
and its global displacement without any parasite reflections at the boundaries  $x=\pm1$}\label{fgwp}
\end{figure}

In figure (\ref{fgwp}) are displayed the density probabilities each 4 time steps.
The time $\tilde{T}$ is equal to $4$, the number of time steps is $N=40$ and $N_x=201$. 
The wave packet velocity $v$ is chosen such that during the time
$\tilde{T}$ a classical particle would travel half the interval $[-1,+1]$.
The initial state is shown by dashed line.
One can see by simple inspection the absence of any parasite reflection nor anomalous behavior.
The curve corresponding to the time $\tilde{t}=\tilde{T}$ is compared to the analytical result.
The difference is not visible by eyes except at the vicinity of $x=1$
in which it reaches its maximum value, less than 1\%.
This difference has been arbitrarily reduced by increasing
the number of time and space grid points showing the validity of this approach.
The parameters chosen in this simple example may serve to illustrate
the efficiency and accuracy of this method.

In the example considered above and due to the value of $v$, the probability flux
leaves the interaction domain mostly at $x=1$.
However, due to the spreading of the wavepacket, the wave function at $x=-1$
is also being populated although its value remains very small and is not visible in the figure.
By taking v=0 one has a symmetric situation for which
the same quality of results has been obtained.

A last check has been done concerning the total probability conservation.
The contribution coming from the interior domain has been obtained
by integrating (trapezoidal rule) the calculated wavefunction.
The contribution from the external part has been evaluated
by equation (\ref{flux1}) which becomes in that case
\begin{equation}\label{flux2}
\int_{1}^{+\infty}\left(\mid\Psi_n(x)\mid^2-\mid\Psi_{n-1}(x)\mid^2\right) dx=
\frac{1}{2\mu^2} (\Psi_n + \Psi_{n-1})^* \partial_x (\Psi_n + \Psi_{n-1}) \mid_{x=1} 
\end{equation}
In this example the probability is found to be conserved better than $10^{-5}$.

\subsection{Scattering by a static potential}

In the second example, the initial wave packet (\ref{gwp}) scatters on an attractive potential:
\[ V(x)= V_0 e^{-{x^2\over b^2} } \]
with $b=0.05$ and $V_0=-150$.
The wavepacket parameters are $x_0=-0.3$, $\sigma_0=0.15$, $v=0.37$. 

\begin{figure}[hbtp]
\begin{center}\epsfxsize=10cm\mbox{\epsffile{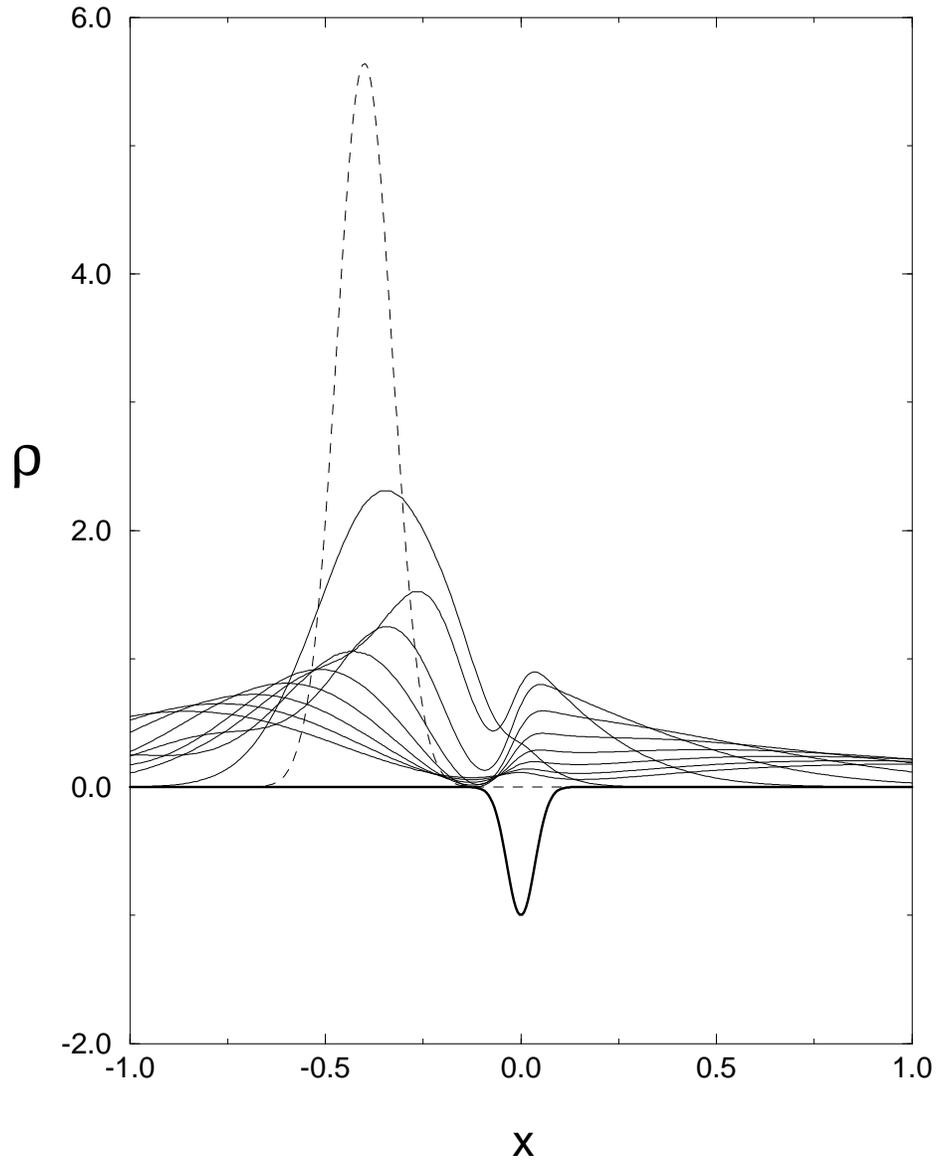}}\end{center}

\vspace{+3.5cm}
\caption{Scattering of a gaussian wavepacket on an attractive gaussian potential
drawn in bold line. The initial state, in dashed line, is centered at $x_0=-0.3$ and has $\sigma_0=0.15$ and $v=0.37$.
It splits into a transmitted and a reflected part which
goes accross the boundaries of the integration domain without any reflection. }\label{static1}
\end{figure}

\begin{figure}[hbtp]
\begin{center}\epsfxsize=10cm\mbox{\epsffile{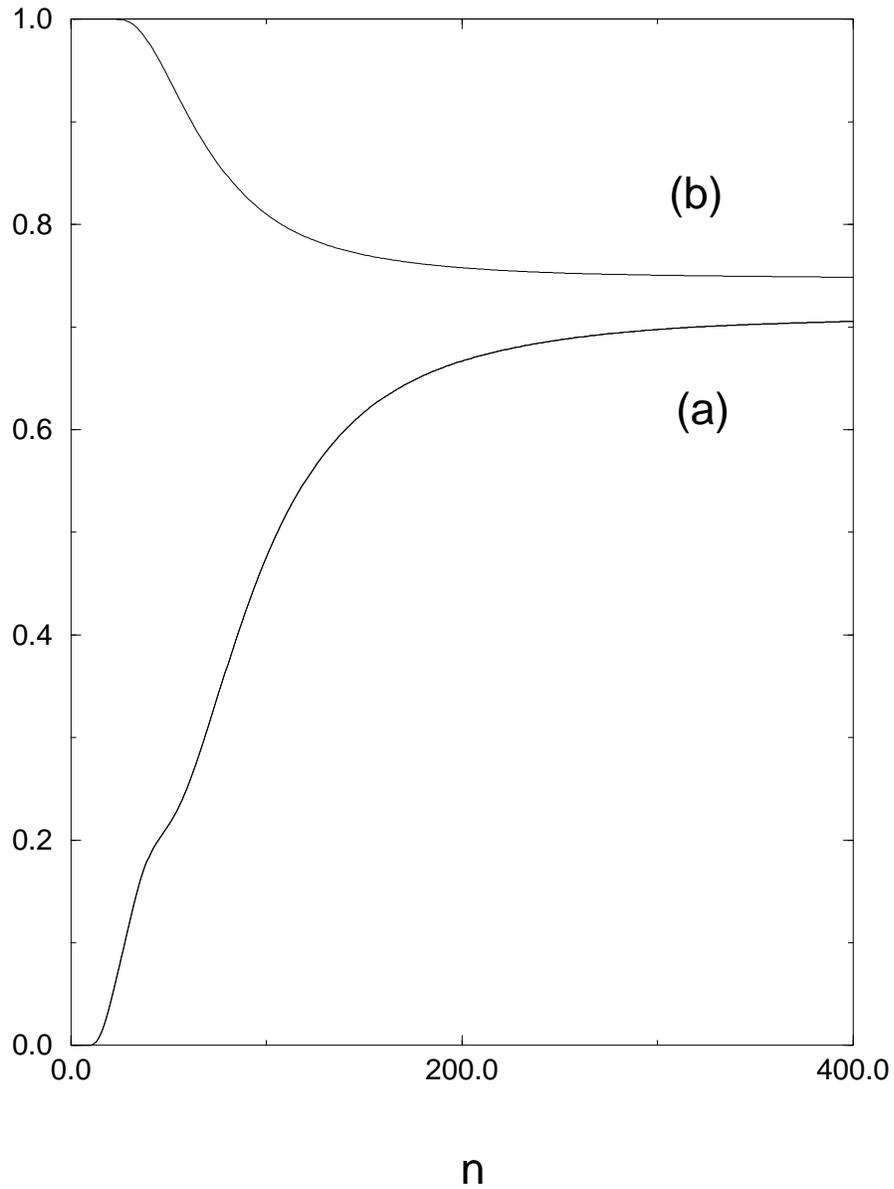}}\end{center}

\vspace{+3.5cm}
\caption{Time evolution of the integrated probability densities for the scattered wavepacket of figure
(\protect\ref{static1}).
Curves (a) and (b) represent the probability density in the intervals $x<-1$ 
and $x<+1$, calculated by equation (\protect{\ref{flux2}}).
They tend towards the corresponding transmission and reflection coefficients.
This convergence is relatively slow due to the slow spreading of the initial state
in the integration domain $D_i=[-1,+1]$.}\label{static2}
\end{figure}

The probability densities at different times are displayed in figure (\ref{static1}).
Its initial value is plotted in dashed line and the attractive well in bold.
The reflected and transmitted wave packets are clearly separated.
The integrated probabilites in $D_e$ and $D_i$ have been calculated
respectively by (\ref{flux2}) and trapezoidal rule from  the solution in $D_i$.
They are shown as a function of time in
figure (\ref{static2}).  The contribution from the $x<-1$ and  $x<1$
regions are respectively plotted in curves (a) and (b).
They correspond to the reflected and transmitted wavepackets.
The difference between both curves corresponds to the
integrated probability in the integration domain. It tends to zero very slowly.

\subsection{Localized state under a time-dependent perturbation}

The third example concerns the evolution of a wave
packet initially localized on an attractive gaussian potential whose amplitude varies periodically:
\begin{equation}\label{TDGP}
 V(x,t)= V_0\left(1+\sin2\pi\omega\tilde{t}\right)\; e^{-{x^2\over b^2}}
\end{equation}
with $b=0.05$, $\omega=0.05$ and $ V_0=-200 $.

\begin{figure}[hbtp]
\begin{center}\epsfxsize=10cm\mbox{\epsffile{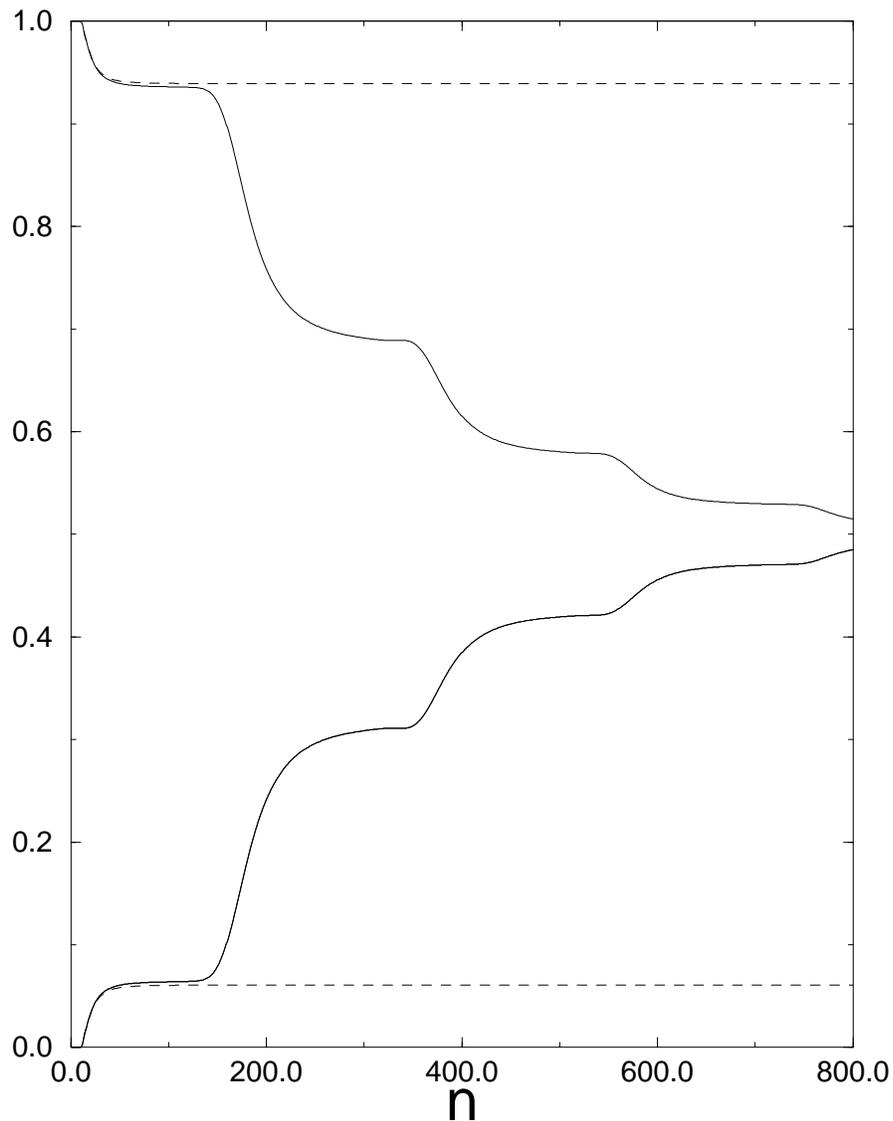}}\end{center}

\vspace{+3.5cm}
\caption{Evolution of a localized state in the time-dependent potential (\protect\ref{TDGP}).
The initial state, a gaussian wavepacket with $\sigma_0=0.10, x_0=0, v=0$, is a superposition of bound states 
plus a small contribution from the continuum, both in the corresponding static potential at $t=0$. 
Curves (a) and (b) represent respectively the probability density in the intervals $x<-1$ and $x<1$.
The difference between these curves, i.e.
 the probability density in the integration domain, tends to zero showing that
the bound state is pulled out of the potential by the time dependent perturbation. 
The evolution in the correponding static potential is shown in dashed lines. }\label{g_t}
\end{figure}

The time of evolution is T=80, with 800 points in space and 800 time-steps.
The initial state is a superposition of the bound states plus a small
component ($ \approx 3 \% $) of continuum states both in the corresponding static potential at $t=0$.
Curves (a) and (b) on figure (\ref{g_t}) represent respectively the probability density
in the intervals $x<-1$ and $x<1$.
The difference between these curves, i.e.
the probability density in the integration domain, tends to zero showing that
the bound state is pulled out of the potential by the time dependent perturbation. 
The evolution in the corresponding static potential, shown in dashed lines, tends to
a constant: the projection of the initial wavepacket into bound states.

\subsection{Tunneling}
 
\begin{figure}[hbtp]
\begin{center}\epsfxsize=10cm\mbox{\epsffile{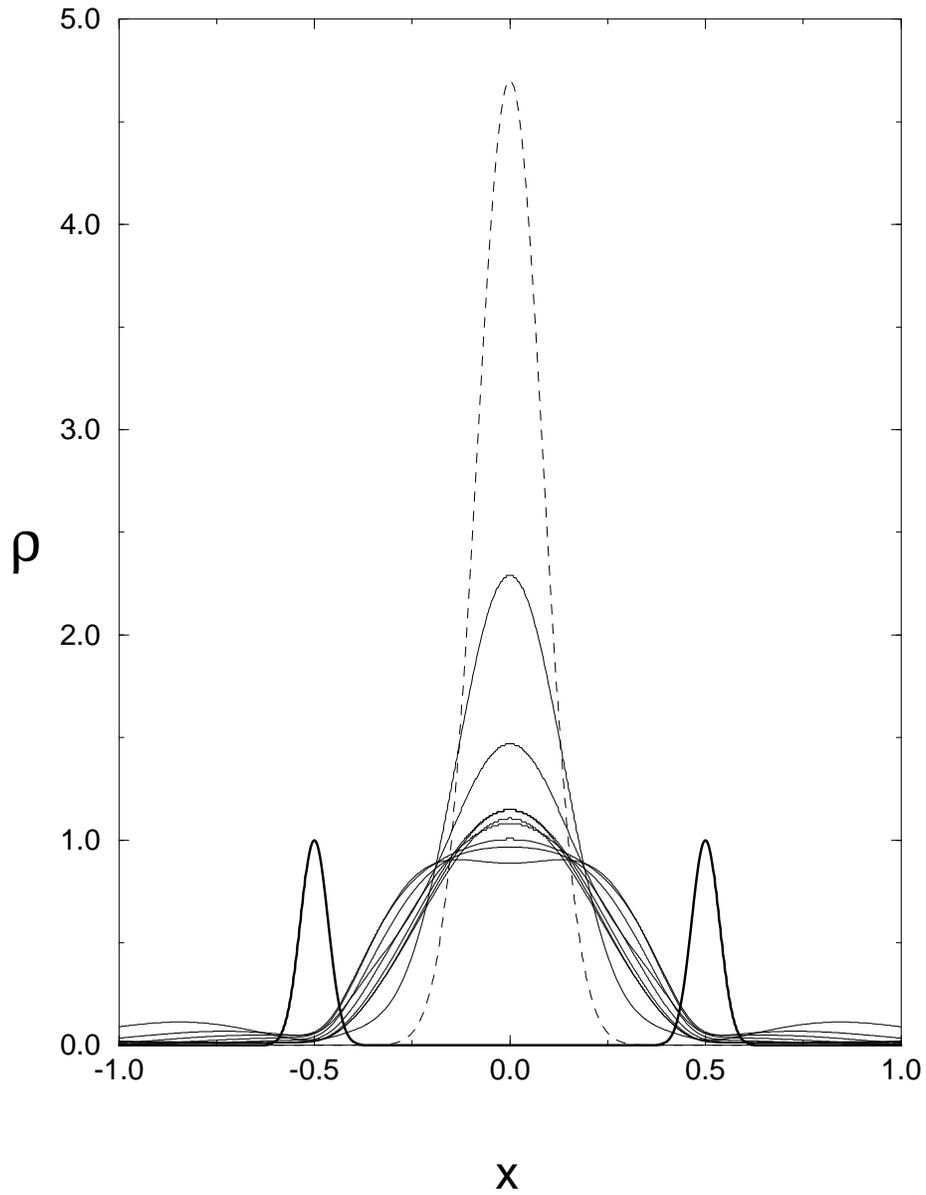}}\end{center}

\vspace{+3.5cm}

\caption{Tunneling through a double repulsive well (bold curve).
The wavepacket is initially centered (dashed line) with $\sigma_0=0.12$ and $v=0$. It spreads
and oscillates inside the potential except for a small part that tunnels through}
\label{tunnel}
\end{figure}

\begin{figure}[hbtp]
\begin{center}\epsfxsize=10cm\vspace{-25.cm}\mbox{\epsffile{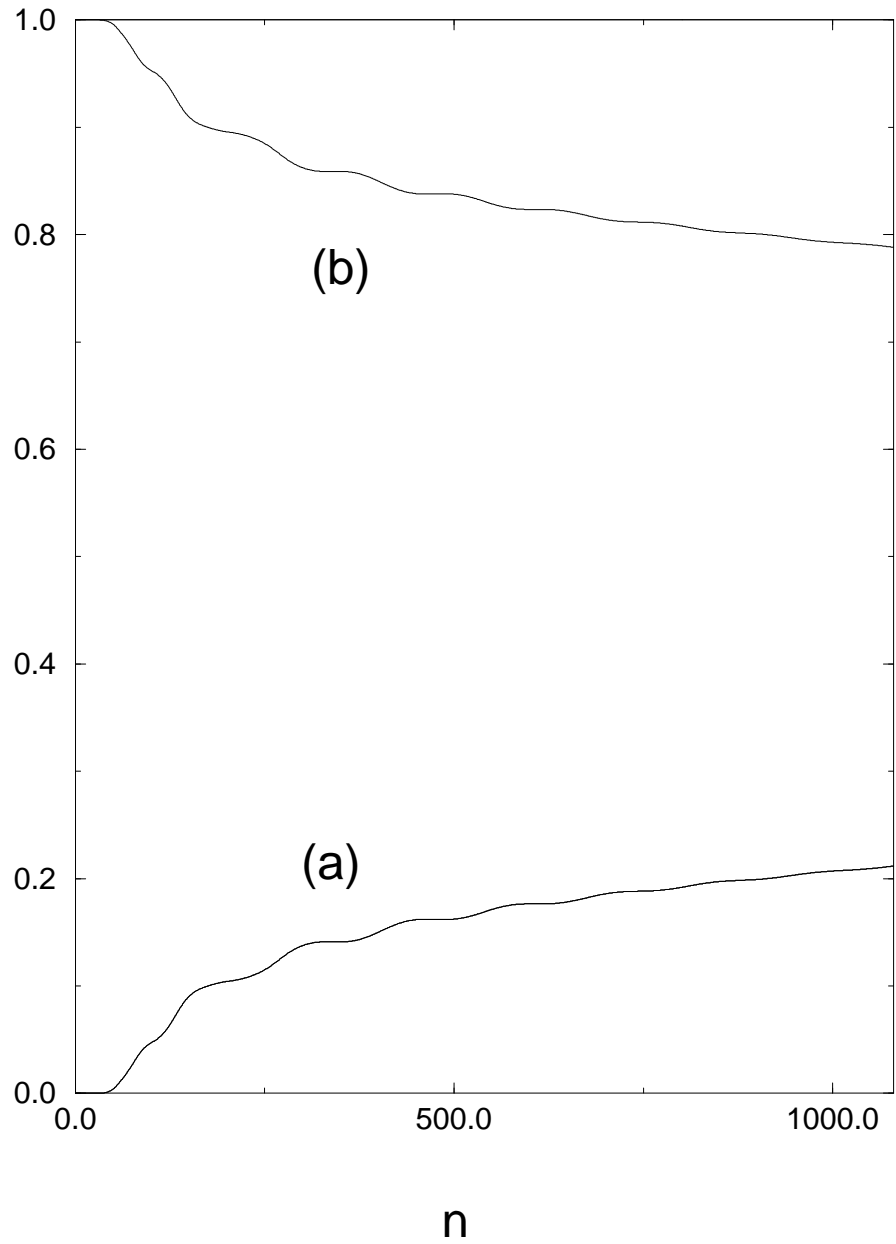}}\end{center}

\vspace{+3.5cm}

\caption{Time evolution of the integrated probability densities corresponding
to Fig. (\protect{\ref{tunnel}}) are shown. Curve (a) is the integrated probability in the region $x<-1$ and curve (b)
in $x<1$. The distance between both curves corresponds
to the integrated probability in the integration domain $D_i$. The small oscillations
are due to the back and forth reflections of the wavepacket inside the well.}
\label{tunnel_p}
\end{figure}

The following example illustrates the tunnel effect through a double well.
A gaussian wavepacket with $\sigma_0=0.12$, initially 
centered and at rest $v=0$ evolves in a double repulsive well of the form 
\begin{equation}\label{dwell}
 V=V_0 \left[ e^{-{(x-a_0)^2 \over b^2}}+e^{-{(x+a_0)^2 \over b^2}} \right]
\end{equation}
with $b=0.05$ $a_0=0.5$  and $V_0=150$. 
We have displayed in figure (\ref{tunnel}) the probability density at different times.
The initial state is drawn in dashed line and the potential (arbitrary units) in bold.
The wavefunction spreads and oscillates inside the potential except for a small part that tunnels through.
The time evolution of the integrated probability densities corresponding
to figure (\ref{tunnel}) is shown in figure (\ref{tunnel_p}).
Curve (a) corresponds to the domain $x<-1$ and curve (b) to $x<1$.
The distance between both curves represents
the integrated probability in the integration domain $D_i$. The small oscillations
correspond to the back and forth reflections of the wavepacket inside the well.

\subsection{Time-dependent Delta potential}\label{delta}

An interesting example is provided by the limiting case of a time-dependent delta potential
\[ V(x)=-\lambda(t)\delta(x)\]
In this case the interaction domain is reduced to a point
and it is possible to obtain closed analytical solutions
for the discrete-time evolution problems.

We will consider the evolution of a state $\Psi$ which initially coincides with
the bound state $\Phi_0$ of the potential with strength $\lambda_0$.
Its energy, in current units ($\hbar=2m=1$), is $E_0=-\omega_0=-{\lambda_0^2\over4}$
and the normalized wavefunction  
\[ \Phi_0(x)=\sqrt{\lambda_0\over2}\; e^{-{\lambda_0\over2}|x|}\]

By integrating the discrete-time Schr\"odinger equation (\ref{DSE}) on both sides of the singularity
one  gets the discontinuity of the wave function derivative at $x=0$
\begin{equation}  \label{delta1}
\lambda_n\left\{\Psi_n(0)+\Psi_{n-1}(0)\right\})+
\left[\partial_x(\Psi_{n}+\Psi_{n-1})\right]_{0^-}^{0^+}=0
\end{equation}
where $\lambda_n$ is the mean intensity of the potential between the $n$ and $n-1$ time-steps.

The solution of (\ref{DSE}) $\Psi_n$ will be written by splitting its 
non perturbed $\Phi_n$ and perturbed $\chi_n$ parts
\begin{equation}\label{phi+chi}
\Psi_n=\Phi_n +\chi_n
\end{equation} 
By definition, $\Phi_n$ is a stationary solution of equation (\ref{DSE}) with
$\lambda=\lambda_0$.
In the continuum case the time evolution of $\Phi_n$ would be simply given by
\[  \Phi_n=e^{-ni\omega_0\Delta t}\Phi_0\]
In the  discrete-time evolution it is easy to show that
\[ \Phi_n= e^{-ni\theta} \Phi_0\]
with
\[ e^{-i\theta}={\mu^2-\omega_0\over\mu^2+\omega_0} \]

The perturbated part $\chi_n$ is a solution of (\ref{FDSE}) on both sides of $x=0$ and
satisfies the boundary conditions (\ref{C1D}). The sum of these two conditions gives
\begin{equation}  \label{delta2}
2\chi_n(0)=-\frac{i}{\mu}\sum\limits_{q=0}^{{n-1 \over 2}}C_q
\left[\partial_x(\chi_{n-2q}+\chi_{n-2q-1})\right]_{0^-}^{0^+}
\end{equation}
Combining this last result with equations (\ref{delta1}) and (\ref{phi+chi}) one obtains at x=0
\begin{equation}\label{delta3}
2\chi_n = \frac{i}{\mu}\sum\limits_{q=0}^{{n-1 \over 2}}C_q
\left[\lambda_n(\chi_{n-2q}+\chi_{n-2q-1})+(\lambda_n-\lambda_0)(\Phi_{n-2q}+\Phi_{n-2q-1})\right]
\end{equation}

This is an inhomogeneous linear system for the perturbed wave function $\chi_n$ with the inhomogeneous
term given by $\Phi_{n}$. By isolating the q=0 contribution equation 
(\ref{delta3}) results into an explicit recurrence relation
\begin{equation} \label{delta5}
(-2i\mu-\lambda_n)\chi_n =\lambda_n\chi_{n-1}
+\sum\limits_{q=1}^{{n-1\over2}}C_q \lambda_n(\chi_{n-2q}+\chi_{n-2q-1})
+\sum\limits_{q=0}^{{n-1\over2}}C_q (\lambda_n-\lambda_0)(\Phi_{n-2q}+\Phi_{n-2q-1})
\end{equation}
The perturbed wave function $\chi_n$ at the origin can be thus easily obtained.
The values at $x\neq0$ can be calculated from equations (\ref{CN5}) and (\ref{Kdx})
This example illustrates how we can extract all the physical
quantities from the knowledge of the wave function and its derivative on the boundary,
even in a limiting case when it is reduced to a point.

\begin{figure}[hbtp]

\begin{center}\epsfxsize=10cm\vspace{-25.cm}\mbox{\epsffile{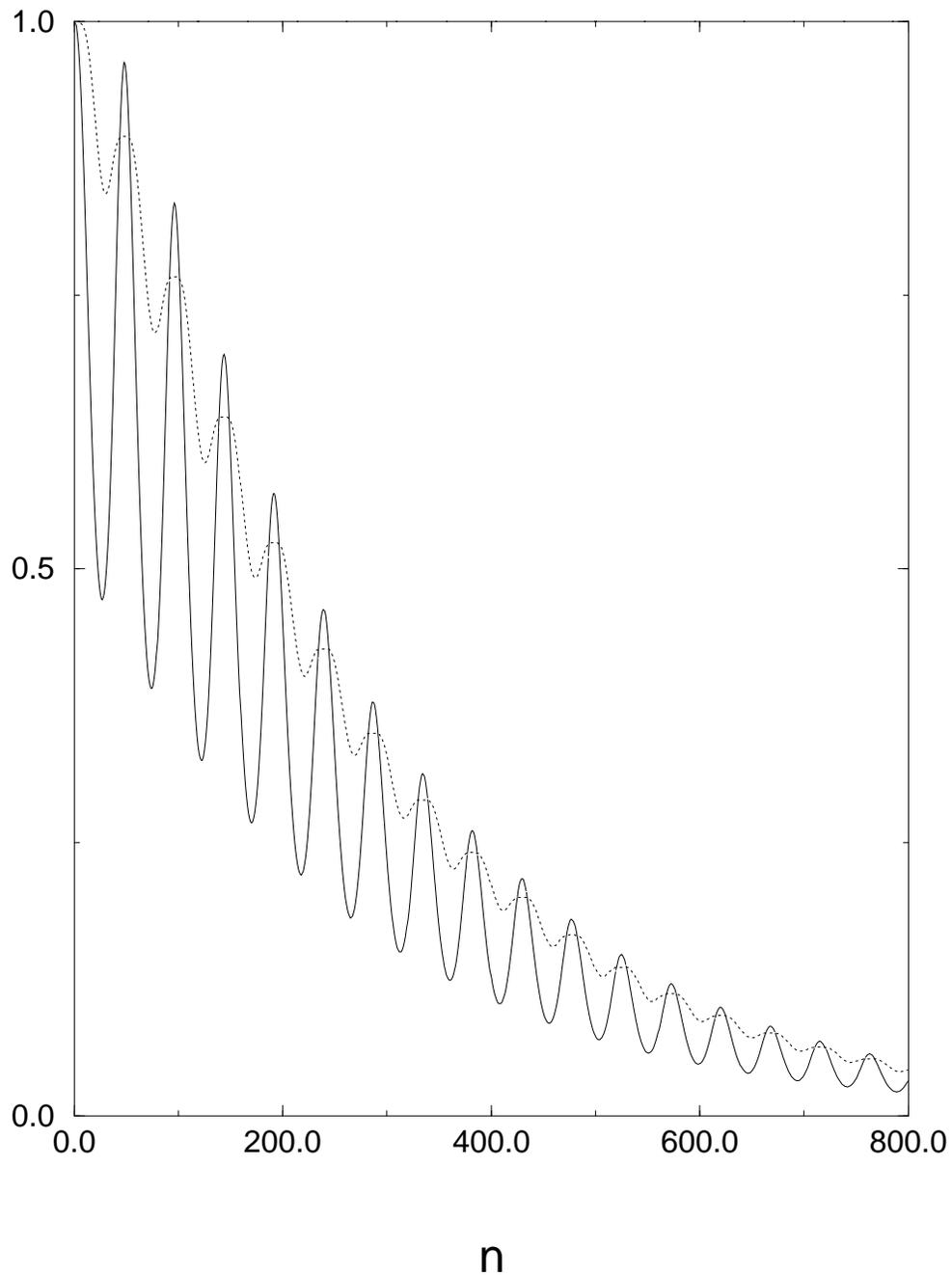}}\end{center}

\vspace{+3.5cm}

\caption{Time-evolution of a bound state in a pulsed delta potential $V(x)=-\lambda(t)\delta(x)$.
Solid curve represents the probability density at the origin.
Dashed curve represents the squared modulus of the autocorrelation function $|{\left\langle\left.{\Psi_0} 
\right| {\Psi_n} \right\rangle}|^2$.
Abscisse units are the number of time-steps. The total time evolution corresponds to 32 Bohr periods. 
The frequency of the  potential pulse is 7/10 the Bohr frequency.}\label{figdel}
\end{figure}

In figure (\ref{figdel}) are displayed the results 
corresponding to a periodically driven delta potential with
\[\lambda(t)=\lambda_0+{A\over2}\left\{1-\cos\left({0.7\omega_0 t}\right) \right\}\]
and where $\omega_0$ is the Bohr frequency of the bound state for $\lambda=\lambda_0$. 
Curve (a) shows the square modulus of the auto-correlation function $<\Phi(t)|\Psi(t)>$
and curve (b) the square modulus of the wave function at the origin normalized to one at the
initial time. For the long times the wave function near the origin becomes
proportional to the bound state and these two quantites are close to each other.
There are 1000 time steps for 40 perturbative pulses.
We note that after a transient period, the wave function is near of an eigenvector 
of the Floquet operator \cite{DH 65} with complex energy.

The periodically driven delta potential is used as a model of weakly bound atom (1D delta atom) 
when studying the tunnel effect in a static exterior field \cite{Gel}.
In this case the exact propagator 
is known explicitly. In our case there is also an explicit solution.

\section{A two dimensional example}

We present in this section a straightforward application of the formulated 
boundary conditions for a two dimensional problem.
We will illustrate it by considering the free evolution of a two dimensional gaussian wave packet
crossing obliquely the $x=1$ axis.
The wave packet has 
initial widths $\sigma_{0x}=\sigma_{0y}=0.2$ and is initially centered at the point $(0,1)$ 
of the XY plane with velocity $v_y/v_x=3/2$.
We push to the infinity the domain for the y variable
and consider as integration domain the band $D_i=[-1,+1]\times {\bf R}$

Let us apply the boundary conditions (\ref{DOBC})  to a point $x_0$ in the line $\gamma_a=\{(a,y)\in
R^2:y\in(-\infty,+\infty)\}$.
The integral term has only contributions coming from $\gamma_a$ itself
and is thus reduced to a one- dimensional integral
although with a two- dimensional propagator of non-vanishing argument  $K_p(\vec{X})$.
\begin{equation}\label{22p}
 {1\over2} \Psi_n(a,y_0)={1\over2\mu^2}\sum_{p=0}^{n-1}\int_{-\infty}^{+\infty} dy \;
\left[  \left( K_{p}(\vec{x}-\vec{x}_0)+K_{p+1}(\vec{x}-\vec{x}_0) \right) 
\nabla  \left( \Psi_{n-p}(\vec{x})+ \Psi_{n-p-1}(\vec{x})          \right) \right]
\end{equation}
with $\vec{x}=(a,y)$ and $\vec{x}_0=(a,y_0)$.

An efficient way to solve (\ref{22p}) is by
Fourier transforming the wavefunction with respect to the $y$ variable
\[ \Psi_n(a,y)=\int_{-\infty}^{+\infty}dk_y \;e^{ik_y y} \; \tilde\Psi_n(a,k_y) \]
and impose the desired boundary condition to each of its Fourier components $\tilde\Psi_n(a,k_y)$.
The problem is then formally equivalent to a one- dimensional one
with a propagator given by expression (\ref{K+K1}) where the $k^2$ variable is replaced by $k^2+k_y^2$.
The resulting integral is no longer analytic as in the one dimensional case but
can be easily evaluated by means of a discrete Fourier tranform according to equation (\ref{Kdx}).

\begin{figure}[hbtp]

\begin{center}\epsfxsize=12cm\vspace{-0.cm}\mbox{\epsffile{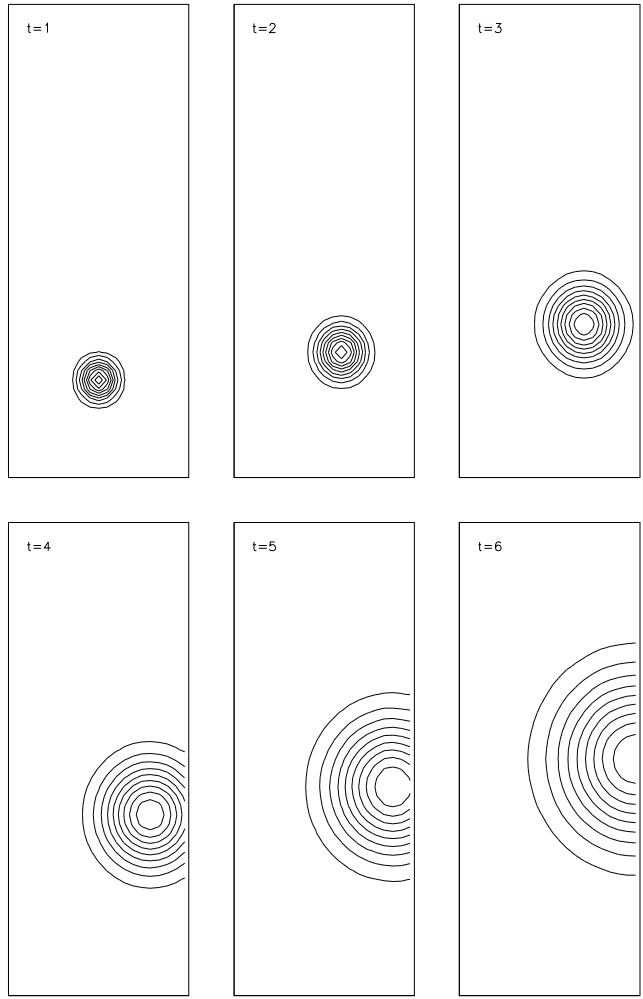}}\end{center}

\vspace{+2.5cm}

\caption{Free evolution of a two-dimensional gaussian wave packet. 
It has been obtained by solving the Schr\"odinger equation in the 
displayed rectangular region, i.e. horizontal axis  $x\in[-1,+1]$ 
and vertical axis  $y\in[0,5]$, by imposing exact boundary conditions at finite distance.
The countour plots of the probability density are shown for six equidistant times. 
The wave packet (t=1) was initially centered at $x=0$ $y=1$ with an initial momentum
$v_y/v_x=3/2$.
The figure shows the spreading of the wave packet in both directions and
the crossing of the $x=1$ boundary without any parasite reflections.}\label{gwp2d}
\end{figure}

The numerical results are displayed in figure \ref{gwp2d}.
They correspond to the solution of the discrete time Schr\"odinger equation in the
region  $x\in[-1,+1]$  and  $y\in[0,5]$ with $N=100$ time steps.
The number of points in x direction is $N_x=101$
and $N_y=45$ points were required for an accurate
Fourier transform of $\Psi_n(a,y)$ amplitudes.
The countour plots of the probability density are shown for six equidistant times. 

The figure shows the spreading of the wave packet in both directions and
the crossing of the $x=1$ boundary without any parasit reflections. 
The comparison of calculated values with the exact analytic expressions
showed a relative accuracy at the maximum of the wave packet better than $1\%$.

\section{Conclusion}

Exact boundary conditions at finite distance for the time-dependent Schr\"odinger equation have been derived.
These conditions allow a resolution of time-dependent quantum mechanical problems
in a finite spatial domain containing the interaction without introducing any parasite reflections,
nor changing the dynamics near the boundary.
They are specially useful to observe long time evolution of waves packets in a finite space domain.

The value of the wavefunction outside the integration domain can be easily obtained from the
values of the wave function and its normal derivative at the boundaries for all the anterior times.

These conditions have been explicitely reformulated in a time-discrete dynamics
and implemented in a numerical algorithm
providing a numerical solution of the more usual one-dimensional problems.
They result in solving  at each discrete time step a stationary Schr\"odinger equation
with complex energy in a finite spatial domain.
The local character of this equation is preserved by the boundary conditions
and we are thus let with a
band matrix problem whatever the discretisation algorithm used.

The economy resulting when solving the problem in a reduced spatial domain
allows to considerably increase the number of grid points and to
use for a same accuracy less powerful but simpler algorithms.  
The numerical results obtained with the Cranck Nicholson scheme
are found to be very satisfactory.
Other time and space discretisation algorithms can however be used as well
at the cost of the simplicity.

The derived boundary conditions
can be applied to more general situations than those considered in the examples.
The case of higher dimensionality and/or of coupled equations, for instance,
can be treated without modifying the formalism.
A numerical application of the free evolution of a two-dimensional gaussian wave packet is given.
Long range static potentials may be taken into account as far as an analytic solution is known
for the corresponding propagators. 
The generalization to the time-dependent treatment of the three body problem is straightforward and
could be of some interest in some molecular physics problems
\cite{LLYM 94} or a doorway to solve the still open question of Coulomb problem above threshold.

The interest in formulating exact boundary conditions at finite distance for 
time-dependent problems recently appeared in the field of statistical
mechanics and conclusions similar to those presented in this work have been
reached in \cite{HF 94}. But this formulation is also relevant in other branches of physics
than those governed by the Schr\"odinger equation.
In particular for the diffusion equation, for which the same results hold with imaginary time.
The case of hyperbolic equation of interest in many
fields of physics can be treated in the same footing \cite{Mep}.

\section{Acknowledgments}
We acknowlegde Dr. K. Protassov for his support on some mathematical aspects of this
work and Prof. C. Leforestier for helpful discussions. 

\bigskip


\newpage

\newpage

\newpage

\newpage

\newpage

\newpage

\newpage

\newpage

\newpage

\end{document}